\documentclass[twocolumn,showpacs,prl]{revtex4}
\usepackage{amstext}
\usepackage{amsmath}            
\usepackage{amssymb}            
\usepackage{graphicx}           
\usepackage{latexsym}
\newcommand{\ket}[1]{\vert#1\rangle}
\newcommand{\bra}[1]{\langle#1\vert}

\begin{document}
\title{Vacuum induced Spin-1/2 Berry phase}
\author{I. Fuentes-Guridi$^{\diamond}$, A. Carollo$^{\diamond\ddag}$, S. Bose$^{\dagger}$  and  V. Vedral$^{\diamond}$}
\address{$^{\diamond}$Optics Section, The Blackett Laboratory,
Imperial College, London SW7 2BZ, United Kingdom \\
$^{\dagger}$ Centre for Quantum Computation, Clarendon Laboratory,
    University of Oxford,
    Parks Road,
    Oxford OX1 3PU, England\\
$^\ddag$INFM, unit\`{a} di ricerca di Palermo, via Archirafi 36,
90123 Italy}

\begin{abstract}
We calculate the Berry phase of a spin-1/2 particle in a magnetic
field considering the quantum nature of the field. The phase
reduces to the standard Berry phase in the semiclassical limit and
eigenstate of the particle acquires a phase in the vacuum. We
also show how to generate a vacuum induced Berry phase considering
two quantized modes of the field which has a interesting physical
interpretation.

\end{abstract}

\maketitle

 Geometric phases in quantum theory attracted great interest since Berry \cite{berry} showed that
 the state of a quantum system acquires a purely geometric feature (called the
Berry phase) in addition to the usual dynamical phase when it is varied slowly and eventually brought back to its
initial form. The Berry phase has been extensively studied \cite{wil,oth}, generalized in various
directions~\cite{aa}, and has very interesting applications, such as implementation of quantum computation by
geometrical means \cite{jvec,ekert00,fal}. In a strict sense, however, the Berry phase has only been studied in a
\emph{semiclassical} context till date. This means that the geometric evolution of a quantum system is studied
under the effect of a time varying classical field. However, this {\em field itself has never been
quantized}~\cite{ken}. Thus, the effects of the vacuum field on the geometric evolution are unknown. Many effects
in quantum optics such as quantum jumps, collapses and revivals of the Rabi oscillations \cite{collapse}, can only
be explained by considering a quantum field, showing the importance of field quantization in the complete
description of physical systems. Moreover, in quantum mechanics several interesting effects are observed due to
the interaction of quantum systems with the vacuum (spontaneous emission, lamb shift)\cite{cat}.

 The canonical experiment that demonstrates the existence of
the Berry phase involves a spin-1/2 particle interacting with an
external magnetic field whose direction is slowly changed in a
cyclic fashion \cite{jvec,nmr}. In this letter we analyze such
experiment in a fully quantized context and give an expression for
the geometric phase of a joint state of the particle and field
that, as expected, reduces to the standard Berry phase in the high
amplitude limit of a coherent state of the field. The relevant
differences between this phase and the semiclassical version of
the Berry phase become evident when states with low photon number
are considered. We show that in the fully quantized scheme it is
possible to produce non-trivial geometric phases that have no
correspondence to phases that can be generated in the
semiclassical scenario. The main difference arises from the
interaction of the spin-1/2 system and the vacuum field. In fact,
we show that even if the field is in the vacuum state, an
adiabatic evolution of the field can be engineered to induce a
non-trivial geometric phase in the system.

We calculate the deviations from the semiclassical
model when a spin-1/2 particle interacts with a single mode
quantum field. In addition we design a scheme to generate a vacuum induced
phase considering the interaction of the particle with two modes
of the field in the vacuum state which are mixed adiabatically to
generate the phase. This phase has a very interesting physical
interpretation which we discuss in this letter. Using a scheme
designed to detect the Berry phase of the joint state of a
harmonic oscillator and a two level system using an ion trap
\cite{us}, it is possible to measure this vacuum induced phase.

  In the semiclassical scenario we consider a spin 1/2 particle,
or more generally, a two level system, coupled to a external
classical oscillating field with frequency $\nu$ not far from the
Bohr frequency $\omega$ of the two level system. In this case, it
is convenient to work in a frame of reference rotating with
frequency $\nu$. The two level system is described in terms of
Pauli operators $\sigma_z$, $\sigma_{\pm}=(\sigma_{x}\pm
i\sigma_{y})/2$ and its dynamics is characterized by the following
Hamiltonian:
\begin{equation}\label{eq:hsemi}
H=\frac{\Delta}{2}\sigma_z + \lambda(\sigma_+ \alpha e^{-i\phi}
+ \sigma_- \alpha e^{i\phi})
\end{equation}
where $\Delta=\omega-\nu$ is the detuning, $\lambda$ is the
coupling constant between the system and the field, and $\alpha$
represents the amplitude of the oscillating field. This
Hamiltonian can be expressed in terms of an effective vector field
${\mathbf B}=(\lambda \alpha\cos\phi,\lambda \alpha\sin\phi,
\Delta/2)$ as $H=\mathbf{B \cdot \sigma}$ where
$\mathbf{\sigma}=(\sigma_x,\sigma_y,\sigma_z)$. When the state of
the two level system is initially prepared in an eigenstate of the
Hamiltonian and the direction of the effective field ${\mathbf B}$
is changed adiabatically, the state of the system will follow the
field and after a complete cycle it will acquire a geometric phase
equal to $\gamma=\pm\frac{1}{2}\Omega$. The $\pm$ sign depends on
whether the state was initially aligned or against the direction
of the field and $\Omega$ is the solid angle subtended by the path
followed by ${\mathbf B}$ in parameter space.

In the simplest case, where the detuning and strength of the
coupling are fixed and the phase $\phi$ varies from $0$ to $2\pi$,
the eigenstates of the particle describe a loop $C$ in the bloch
sphere and Berry's adiabatic geometric phase is (up to a $\pm$
sign)
\begin{equation} \label{eq:gamma}
\gamma=\pi(1-\cos{\theta}).
\end{equation}
where $\cos{\theta}=\Delta/\sqrt{\Delta^{2}+4(\alpha\lambda)^{2}}$.

We will now calculate the phase taking into account that a more
rigorous description of the oscillating field that drives the
spin-1/2 particle involves a quantum field. This means that the
field can not be considered anymore an external variable, it is
part of the system itself, and its state can be manipulated by
varying the parameters of the Hamiltonian. In the initial
situation we consider the Hamiltonian of a spin-1/2 particle
interacting with a single quantized mode of the field in the
rotating wave approximation \cite{jcm}
\begin{equation} \label{eq:hone}
H^{q}_{o}=\nu
a^{\dagger}a+\frac{\omega}{2}\sigma_{z}+\lambda(\sigma_{+}a+\sigma_{-}a^{\dagger}),
\end{equation}
where $\nu$ is the frequency of the field described in terms of
the creation and annihilation operators $a$ and $a^{\dagger}$,
$\omega$ is the transition frequency between the eigenstates of
the particle and $\lambda$ the coupling constant. It can be seen
by replacing the operators $a$ and $a^{\dagger}$ by the classical
amplitude $\alpha$ that this Hamiltonian corresponds to the
semiclassical Hamiltonian (\ref{eq:hsemi}) before the rotation
involving $\phi$ has been applied.
In the standard semiclassical experiment previously introduced,
the variation of the state was induced by an adiabatic change of
the phase of classical field. In the fully quantized context we
need a procedure capable to generate an analogous phase change in
the state of the field. To this end we introduce the phase shift
operator $U(\phi)=\exp{(-i\phi a^{\dagger}a)}$ that, applied
adiabatically to the Hamiltonian of the system, is capable to
change the state of the field. Changing $\phi$ slowly from $0$ to
$2\pi$ the Berry phase generated is calculated as follows
\begin{equation}
\gamma^{q}_{\pm}=i\int_{c}
d\phi\langle\Phi_{n}^{\pm}|U(\phi)^{\dagger}\frac{d}{d\phi}U(\phi)|\Phi_{n}^{\pm}\rangle.
\end{equation}
where $|\Phi_{n}^{\pm}\rangle$ are the eigenstates of the Hamiltonian~(\ref{eq:hone}).
Substituting the expression of $|\Phi_{n}^{\pm}\rangle$ leads to
\begin{eqnarray}
\gamma^{q}_{+}&=&\pi(1-\cos{\theta_{n}})+2\pi n,\label{eq:fulqbp1}\\
\gamma^{q}_{-}&=&-\pi(1-\cos{\theta_{n}})+2\pi(n+1).\label{eq:fulqbp2}
\end{eqnarray}
with
$\cos{\theta_{n}}=\Delta/\sqrt{\Delta^{2}+4\lambda^{2}(n+1)}$. It
is important to note that for $n=0$ the phase is different from
zero, which means that the vacuum field introduces a correction in
Berry's phase. Moreover, if we consider a coherent state with
large amplitude we recover the semiclassical result
($\cos\theta_{n}\approx\cos\theta$). We now have a general
expression for the Berry phase considering the quantum nature of
light. This expression is relevant when systems are driven by
fields with few photons and Berry's result is recovered when the
photon number grows.

In order to study the physical meaning of the second term we need a non-trivial contribution that is different from an integer multiple of $2\pi$. Therefore,
we now describe a scheme were the particle interacts with two modes
of the field. Consider the initial Hamiltonian
\begin{equation} \label{eq:twomode}
H^{2q}_{o}=\nu a^{\dagger}a+\nu
b^{\dagger}b+\frac{\nu}{2}\sigma_{z}+\lambda(\sigma_{+}a+\sigma_{-}a^{\dagger}),
\end{equation}
describing a spin-1/2 particle interacting with one mode of the
field with creation and annihilation operators $a$ and
$a^{\dagger}$ through a Jaynes-Cummings resonant interaction with
coupling constant $\lambda$ and a second mode of the field with
creation and annihilation operators $b$ and $b^{\dagger}$ and
frequency $\nu$ which initially does not interact with the
particle nor the first mode of the field. The eigenstates of this
Hamiltonian are
\begin{equation} \label{eq:states2}
|\Psi^{\pm}_{n,n^{\prime}}\rangle=\frac{1}{\sqrt{2}}(|e,n\rangle\pm|g,
n+1 \rangle)|n^{\prime}\rangle.
\end{equation}
The state vector is a product state of the states
$|n^{\prime}\rangle$ of the field with modes $b$ and $b^{\dagger}$
and the Jaynes-Cummings eigenstates of joint state of the field
with modes $a$ and $a^{\dagger}$ and the particle. Now we are
allowed to exploit the second mode to perform a more general class
of transformations, using the mode $b$ as an "ancilla". Instead of
changing phase of the field and detuning between the two level
system and the field, we consider the possibility of linearly
mixing the two modes in a cyclic way, without any direct action on
the degrees of freedom of the two level system. Before considering
this transformation let's introduce some notation. The operation
of linear mixing of two modes can be represented in a suitable way
employing the Schwinger angular momentum ($SU(2)$) operators
\begin{eqnarray}
&J_{z}=\frac{1}{2}(a^{\dagger}a-b^{\dagger}b)\nonumber&\\
&J_{x}=\frac{1}{2}(a^{\dagger}b+ab^{\dagger}),\quad
J_{y}=\frac{1}{2i}(a^{\dagger}b-ab^{\dagger}).&\nonumber
\end{eqnarray}
The physical meaning of this operator can be easily understood if
we look at the two modes as two different polarization of the same
spatial mode. The Schwinger operators can then be interpreted as
the generators of the polarization transformations, or in more
abstract terms they determine rotations in the Poincare's
sphere $\mathcal{S}^2\sim SU(2)/U(1)$.
In order to generate a Berry phase we perform a cyclic loop in the
Poincare's sphere, changing adiabatically the Hamiltonian by means
of the unitary transformation:
\begin{equation} \label{eq:evolv}
U(\theta,\phi)=\exp{(-i\phi J_{z})}\exp{(-i\theta J_{y})},
\end{equation}
where $\theta$ and $\phi$ are slowly varying parameters. The
transformed Hamiltonian is
\begin{eqnarray}
\lefteqn{H^{2q}=UH_{o}U^{\dagger}=\nu(\sigma_{z}/2+a^{\dagger}a+b^{\dagger}b)+}
\\&&{}+\lambda(\cos{\theta/2}\sigma_{+}ae^{-i\phi/2}+\sin{\theta/2}\sigma_{+}be^{i\phi/2}+h.c.).\nonumber
\end{eqnarray}
This Hamiltonian describes spin-1/2 particle interacting
simultaneously with two modes of the field through a
Jaynes-Cummings Hamiltonian where the mode phases and the coupling
between spin-1/2 particle and each modes are adiabatically varied
through the parameters $\theta$ and $\phi$. Both eigenstates
(\ref{eq:states2}) acquire the phase
\begin{equation} \label{eq:fullygamma}
\gamma_{n,n^{\prime}}=\frac{1}{2}\Omega(n-n^{\prime}+1/2),
\end{equation}
where $\Omega$ is the solid angle subtended by the cyclic loop in
the Poincare's sphere. The phase dependence on the photon number
is not trivial and can be measured by using an interference
procedure between any of the eigenstate
$\ket{\Psi^\pm_{n,n^\prime}}$ and the ground state $\ket{g,0,0}$,
which is the only state that acquires no geometric phase. The most
remarkable case is obtained with the initial state
$\ket{e,0,0}=\frac{1}{\sqrt{2}}(\ket{\Psi^{+}_{0,0}}+\ket{\Psi^{-}_{0,0}})$.
Indeed, even though the field is in a vacuum state the geometric
operations performed on the degrees of freedom of the field
determine a Berry's phase
\begin{equation}\label{vacuum:phase}
\gamma_{zero}=\frac{\Omega}{4}.
\end{equation}
Clearly, this result has no semiclassical correspondence, on
account of the absence of a classical interpretation of a vacuum
state. However, it is worth investigating the physical meaning of
the equation~(\ref{eq:fullygamma}). An interpretation of the first
part of equation~(\ref{eq:fullygamma}), namely
$\Omega(n-n^\prime)/2$, can be provided in terms of a polarized
field not-interacting with the two level system, subjected to a
rotation of its polarization~\cite{wil,pac}. Indeed, this term has a classical
origin that can be understood as follows. Suppose a beam of
classical polarized light traverses an anisotropic dielectric such
that its polarization slowly rotates and performs a closed loop in
the Poincare's sphere due to the variation of dielectric
properties along the direction of propagation. According to
Maxwell's equations, if the variation in the medium is slow
enough, the beam of light acquires a geometric phase. Therefore,
if $\alpha$ and $\beta$ are the complex amplitudes of the two
eigenmodes of the dielectric, under the cyclic rotation in the
Poincare's sphere they becomes $\alpha e^{i\Omega/2}$ and $\beta
e^{-i\Omega/2}$, respectively. If the same experiment is analyzed
in a context where the electromagnetic field is quantized, we
expect to see the same effect. If we quantize the two modes by
substituting the complex amplitude $\alpha$ and $\beta$ with the
corresponding annihilation operators $a$ and $b$, the effect of
the geometric evolution on the Fock states of the field is:
\begin{equation}\label{f:noint}
\ket{n,n^\prime}\rightarrow
e^{i(n-n^\prime)\Omega/2}\ket{n,n^\prime}
\end{equation}
Clearly, if we consider coherent states of the field the classical
result is recovered: $\ket{\alpha}\ket{\beta}\to\ket{\alpha
e^{i\Omega}}\ket{\beta e^{-i\Omega}}$. Therefore, if we look at
the two modes $a$ and $b$ in equation~(\ref{eq:twomode}) as two
different polarizations of the same spatial mode,
$\Omega(n-n^\prime)/2$ is the geometric phase corresponding to the
polarization rotation of an electromagnetic field not interacting
with the two level system.
Then the question is: what is the physical origin of the
term~(\ref{vacuum:phase})? Clearly the interaction with the two
level system is responsible for the appearance of this term. To
have a picture of how this term come into play, we can consider
the semiclassical limit of the two level system interacting with
the field. Suppose that modes $a$ and $b$ of the field initially
are in a coherent state $\ket{\alpha}$ and $\ket{\beta}$,
respectively, and the particle is in the  linear combination
$(|e\rangle\pm|g\rangle)$. Now, the state is adiabatically
transformed by means of the operator~(\ref{eq:evolv}) whose
parameter are slowly varied along a close loop in the parameter
space ${\theta,\phi}$. In the limit of large amplitude coherent
state $\ket{\alpha}$ ($|\alpha|>>1$) the state is transformed as
\begin{eqnarray}\label{quant-class}
\lefteqn{(|e\rangle \pm |g\rangle)|\alpha,\beta \rangle \to}\\
&&\to(e^{i\Omega/4}|e\rangle\pm
e^{-i\Omega/4}|g\rangle)|e^{i\Omega/2}\alpha,e^{-i\Omega/2}\beta\rangle.\nonumber
\end{eqnarray}
After the adiabatic evolution, the amplitudes $\alpha$ and $\beta$
of the coherent states acquire a phase $\Omega/2$ and $-\Omega/2$,
respectively. According to the result~(\ref{f:noint}) obtained in
the case of a field non-interacting with the two level system,
this phase is associated with the polarization rotation, and
originates from the term $\Omega(n-n^\prime)/2$ of the
equation~(\ref{eq:fullygamma}). Since in the large amplitude
coherent state the two level system is approximately not entangled
with the field, the term~(\ref{vacuum:phase}) appears in the last
equation as a phase associable to the state of the two level
system only. Under this condition exist a possible explanation of
the phase~(\ref{vacuum:phase}) in terms of a semiclassical model.
Consider the semiclassical Hamiltonian~(\ref{eq:hsemi}) in the
resonant case ($\Delta=0$). Suppose to rotate the polarization
direction of the classical field. Coherently with the previous
notation we describe this polarization with the two dimensional
complex unit vector
\begin{equation}
\mathbf{\epsilon}(\theta,\phi)=
\left(\begin{array}{c}
e^{i\phi/2}\cos{\theta/2}\\e^{-i\phi/2}\sin{\theta/2}
\end{array}\right)=A\left(\begin{array}{c} \alpha\\\beta
\end{array}\right),
\end{equation}
where $A=1/\sqrt{|\alpha|^2+|\beta|^2}$. During the rotation the
field acquires a geometric phase $\gamma(s)$ that depends on the
path $s$ followed by the the vector $\mathbf{\epsilon}$ on the
Poincare's sphere. Since this phase is not integrable, the state
of the field cannot be specified completely in terms of the
parameters $\theta$ and $\phi$ only, but it needs to be expressed
also as a function of the path $s$. Then, in a more complete
description, the state of polarization has to be expressed as $
\tilde{\mathbf{\epsilon}}(\theta,\phi)=e^{i\gamma(s)}\mathbf{\epsilon}(\theta,\phi).
$ Since the field is a parameter of the semiclassical Hamiltonian,
in the expression~(\ref{eq:hsemi}) the field must be considered
together with its phase factor $e^{i\gamma(s)}$. Therefore, the
Hamiltonian is no longer a function of $\theta$ and $\phi$ only,
but depends, through the field, also on its previous "history".
Taking into account this further phase in the
expression~(\ref{eq:hsemi}) leads to
\begin{equation}\label{eq:hsemigp}
H= \lambda A\sigma_+e^{i\phi/2}\cos{\theta/2}e^{i\gamma(s)}+ h.c.\
.
\end{equation}
This means that the eigenstates of~(\ref{eq:hsemigp}) are
functions also of the geometric phase $\gamma(s)$. The presence
of the geometric phase in the Hamiltonian becomes non trivial when
the field performs a closed loop in its parameter phase. At the
end of this loop the state have experienced a global
transformation due to the geometric phase accumulated by the field
in its evolution. For example, starting from parameters $\theta=0$
and $\phi=0$ after a closed loop the Hamiltonian transforms into
$H=\lambda A(\sigma_+e^{i\Omega/2}+h.c.)$ and accordingly, its
eigenstates  transform as
$ |e\rangle\pm|g\rangle\rightarrow e^{i\Omega/4}|e\rangle\pm
e^{-i\Omega/4}|g\rangle,$
which is the result expected from the quantum description
(see~(\ref{quant-class})).

On the other hand the classical picture fails to explain the phase
$\Omega/4$ when the system involve a field with low number of
photons. The entanglement in the eigenstates~(\ref{eq:states2})
cannot be neglected in this case, and the
expression~(\ref{vacuum:phase}) cannot be interpreted as a phase
of the two level system only. A remarkable explanation of the
origin of the phase $\Omega/4$ can be provided in terms of the
vacuum fluctuation of the field. Due to the entanglement of the
eigenstates~(\ref{eq:states2}) the field is not in a pure states,
and must be considered as an incoherent combination of
$\ket{n,n^\prime}$ and $\ket{n+1,n^\prime}$. It is still possible
to provide an operationally well defined generalization of
geometric phase when a system is in a mixed state. According to
the definition introduced by Sj\"oqvist et al.~\cite{sjoq}, the
geometrical phase for a mixed states $\rho=\sum_k w_k
\ket{k}\bra{k}$ evolving under a closed, adiabatic transformation
$U$ can be expressed in terms of an average connection form:
\begin{equation}\label{mixd}
e^{i\gamma}=\sum_k w_k e^{i\gamma_k}
\end{equation}
where
 $\gamma_k$ is Berry's phase related to the
eigenstate $\ket{k}$. Applying this concept to the state of the
field
$\rho=1/2(\ket{n}\bra{n}+\ket{n+1}\bra{n+1})\otimes\ket{n^\prime}\bra{n^\prime}$,
evolving under the transformation~(\ref{eq:evolv}) leads to the
geometric phase given by the expression~(\ref{eq:fullygamma}). In
case of $n=n^\prime=0$ the phase $\Omega/4$ is obtained. This
phase is therefore determined by the vacuum photon fluctuation due
to the interaction with the two level system. The average number
of photons in the the state of the field determines the non
integer number "$n-n^\prime+1/2$" multiplying the classic
geometric phase $\Omega/2$ in the
expression~(\ref{eq:fullygamma}).

The Berry phase is usually described as the phase acquired by the
state of a system when an adiabatic and cyclic change in the state
is generated by means of variations on external classical fields
acting on a quantum system. However, in physics there are many
striking effects of field quantization. We have investigated the canonical example of the Berry
phase giving a rigourous description of the field. We considered the
quantum nature of the field in the geometric evolution for a
spin-1/2 particle rotating in a cyclic fashion by means of a
slowly varying magnetic field and found a general expression for
the phase which reduces to the standard result in the
semiclassical limit and presents vacuum induced effects.
In addition, we have shown how to generate a vacuum induced phase
in the state of a spin-1/2 particle using the most general
rotation in the space of the system, which corresponds to a two
mode displacement operator that mixes the modes of the field. This
result opens up a new arena to study the consequences of field
quantization in the geometric evolution of states. We are
investigating possible applications of this effect and its
connections to other quantum effects in different systems.

We thank J. Anandan and D. Markham for valuable comments and
discussions. This research has been partly supported by the
European Union, EPSRC, Hewlett-Packard and EU under Grant Nos. TMR
ERB FMR XCT 96-0087 and IST 1999-11053-EQUIP. I. F.-G. would like
to thank Consejo Nacional de Ciencia y Tecnologia (Mexico) Grant
no. 115569/135963 for financial support.


\begin{thebibliography}{99}

\bibitem{berry} M. V. Berry, Proc. Roy. Soc. A {\bf 392}, 45 (1984).
%
\bibitem{wil} A. Shapere and F. Wilczek eds., {\em Geometric phases in physics}
(World Scientific, Singapore, 1989).

\bibitem{oth}
D. Thouless {\em et al.}, Phys. Rev. Lett. {\bf 49}, 405 (1983);
F. S. Ham, Phys. Rev. Lett. {\bf 58}, 725 (1987); H. Mathur, Phys.
Rev. Lett. {\bf 67}, 3325 (1991); H. Svensmark and P. Dimon, Phys.
Rev. Lett. {\bf 73}, 3387 (1994); M. Kitano and T. Yabuzaki, Phys.
Lett. A {\bf 142}, 321 (1989).
%
\bibitem{aa}
Y. Aharonov and J. Anandan, Phys. Rev. Lett. {\bf 58}, 1593
(1987); J. Samuel and R. Bhandari, Phys. Rev. Lett. {\bf 60}, 2339
(1988); N. Mukunda and R. Simon, Ann. Phys. (NY), 205 (1993); {\em
ibid}, 269 (1993); A. K. Pati, Phys. Rev. A {\bf 52}, 2576 (1995).
%
\bibitem{jvec} J.A. Jones, V. Vedral, A. Ekert,
and G. Castagnoli, Nature {\bf 403}, 869 (1999).
%
\bibitem{ekert00} A. Ekert, M. Ericsson, P. Hayden,
H. Inamori, J.A. Jones, D.K.L. Oi, and V. Vedral, J. Mod Opt. {\bf 47}, 2051,
(2000).
%
\bibitem{fal} G. Falci, R. Fazio, G. M. Palma, J. Siewert, V. Vedral
Nature {\bf 407}, 355 (2000).
%
\bibitem{ken} The quantisation of a driving external system has been taken into account in the context of molecular physics.
 In that scenario the "slow degrees of freedom" of the molecules play the role of external fields for the "fast degrees of freedom", see for example: C. A. Mead, Rev. Mod. Phys. {\bf 64}, 51 (1992); B. K. Kendrick, J. Chem. Phys. {\bf 112} 5679 (2000).
%
\bibitem{collapse}J. H. Eberly, N. B. Narozhny and J. J. Sanchez-Mondragon, Phys. Rev. Lett. {\bf 44},1323 (1980)%
%
\bibitem{cat} R. Loudon and P. Knight, J. Mod Opt. {\bf 34}, 709,
(1987).
%
\bibitem{nmr} D. Suter, G. C. Chingas, R. A. Harris, A. Pines, Mol. Phys. {\bf 61}, 1327 (1987)
\bibitem{us} I. Fuentes-Guridi, S. Bose and V. Vedral, Phys. Rev. Lett. {\bf 85}, 5018
(2000);
B. E. King, Chapter $9$, {\em PhD thesis}, unpublished (1999).
%
\bibitem{jcm}
B. W. Shore and P. L. Knight, J. Mod. Opt. {\bf 40}, 1195 (1993).
%
\bibitem{pac}
S. Pancharatnam, Proc. Ind. Acad. Sci. A {\bf 44}, 247 (1956); M. V. Berry J. Mod. Opt. {\bf 34}, 1401 (1987);
%
\bibitem{sjoq}
E.\ Sj\"oqvist, A.\ K.\ Pati, A.\ Ekert, J.\ S.\ Anandan, M.\ Ericsson, D.\ K.\ L.\ Oi and V.\ Vedral, Phys.\
Rev.\ Lett. {\bf 85}, 2845 (2000).
\end{thebibliography}
\end{document}